\documentclass{iau} 
\usepackage{graphicx}
\usepackage{natbib}
\usepackage{aas_macros}
\usepackage{array}

\title[Resolving Power of Asteroseismic Inversion] 
{Resolving Power of Asteroseismic Inversion of the Kepler Legacy Sample}

\author[A.G. Kosovichev \& I.N. Kitiashvili]   
{Alexander G. Kosovichev$^1$ \and Irina N. Kitiashvili$^2$}

\affiliation{$^1$New Jersey Institute of Technology, Newark, NJ 07102, U.S.A.
	\\ email: {\tt alexander.g.kosovichev@njit.edu}
\\[\affilskip]
$^2$ NASA Ames Research Center\\ Moffett Field, CA 94035, U.S.A.
	\\ email: {\tt irina.n.kitiashvili@nasa.gov}}

\pubyear{2020}
\volume{354}  
\jname{Solar and Stellar Magnetic Fields: Origins and Manifestations}
\editors{A.G. Kosovichev, K. Strassmeier \& M. Jardine, eds.}

\begin{document}
\maketitle

\begin{abstract}
The Kepler Asteroseismic Legacy Project provided frequencies, separation ratios, error estimates, and covariance matrices for 66 Kepler main sequence targets. Most of the previous analysis of these data was focused on fitting standard stellar models. We present results of direct asteroseismic inversions using the method of optimally localized averages (OLA), which effectively eliminates the surface effects and attempts to resolve the stellar core structure. The inversions are presented for various structure properties, including the density stratification and sound speed. The results show that the mixed modes observed in post-main sequence F-type stars allow us to resolve the stellar core structure and reveal significant deviations from the evolutionary models obtained by the grid-fitting procedure to match the observed oscillation frequencies.
\end{abstract}

\section{Introduction}

The Kepler mission \citep{Borucki2010} provided a wealth of stellar oscillation data enabling asteroseismic investigation of the internal structure and rotation of many stars across the HR diagram. A primary tool of asteroseismology employed for interpretation of observed oscillation frequencies used a method of grids of stellar models. In this method, the asteroseismic calibration of stellar models is performed by matching the observed oscillation frequencies to theoretical mode frequencies calculated for a grid of standard evolutionary models. In combination with spectroscopic constraints, this approach provided estimates of stellar radius, composition, and age with unprecedented precision \citep[see][and referencies therein]{Chaplin2010,Metcalfe2010}. In addition, the high accuracy measurements of oscillation frequencies opened a new opportunity for asteroseismic inversions which allows us to reconstruct the internal structure and test evolutionary stellar models. A similar approach has been used in helioseismology for more than two decades.

A specific feature of asteroseismology data is that only low-degree oscillations can be observed, typically for modes of spherical harmonic degrees, $\ell=0,1,2$, and sometimes, $\ell=3$. In this situation, only the structure of stellar cores can be resolved by inversion techniques. An additional difficulty is caused by uncertainties in the mass and radius of distant stars. \citet{Gough1993} showed that this difficulty can be overcome by an additional condition in the inversion procedure, which constraints the frequency scaling factor, $q=M/R^3$, where $M$ and $R$ are the stellar mass and radius. They showed that if the low-$\ell$ mode frequencies are measured with precision of $0.1~\mu$Hz then the structure of the stellar core can reconstructed even when the stellar mass and radius are not known. When the measurement precision is relatively low (e.g. $1~\mu$Hz), the localization of the inversion accuracy is degraded. For such cases, \citet{Gough1993b} suggested a procedure of calibration of the averaging kernels to estimate averaged properties of the stellar core. The effectiveness of this procedure was demonstrated by applying it to the low-degree solar oscillation frequencies observed by the IPHIR instrument that measured the total solar irradiance onboard the PHOBOS  spacecraft\citep{Toutain1992}. For analysis of the Kepler asteroseismology data, the inversion technique has been recently used by \citet{Bellinger2019}, and the kernel calibration method was applied by \citet{Buldgen2017a} to estimate integrated properties of stellar structure.
\begin{table}
	\begin{center}
		\caption{Characteristics of the stellar models that are used for inversion of the observed oscillation frequencies.}\label{Table1}
		\begin{tabular}{|c|c|c|c|c|c|c|c|c|} \hline 
			KIC & M/M$_\odot$ & lg(R/R$_\odot$) & Age(Gyr) & lg(L/L$_\odot$) & Ysurf & Zsurf & Teff & alpha \\ \hline 
			7206837 & 1.298 & 0.1980 & 2.90 & 0.5523 & 0.2800 & 0.0220 & 6320 & 1.791 \\ \hline 
			1435467 & 1.382 & 0.2436 & 2.56 & 0.6694 & 0.2637 & 0.0197 & 6414 & 1.692 \\ \hline 
			10162436 & 1.461 & 0.3082 & 2.51 & 0.8110 & 0.2460 & 0.0173 & 6494 & 1.684 \\ \hline 
			9353712 & 1.516 & 0.3261 & 2.03 & 0.9008 & 0.2603 & 0.0180 & 6665 & 1.713 \\ \hline 
			5773345 & 1.579 & 0.3074 & 2.07 & 0.7931 & 0.2593 & 0.0306 & 6400 & 1.715 \\ \hline 
			12069127 & 1.588 & 0.3403 & 1.76 & 0.9385 & 0.2669 & 0.0216 & 6700 & 1.672 \\ \hline 
		\end{tabular}
	\end{center}
\end{table}

In this paper, we show that the detection of mixed modes in the oscillation spectra of F-type stars allows us to resolve the structure of the inner stellar cores. The mixed modes have properties of internal gravity waves (g-modes) in the convectively stable helium core and properties of acoustic modes outside the core. The oscillation frequencies of these modes may be quite sensitive to the properties of the core. Including them in the inversion procedure allows us to localize the averaging kernels in the core region. This opens a unique opportunity for testing the stellar evolution theory for subgiant stars where the energy release is in a hydrogen shell surrounding a helium core.
\section{Target selection. Evolutionary models.}
From the Kepler Asteroseismic Legacy Sample \citep{SilvaAguirre2017a} we selected 6 stars in the mass range from about 1.3 to 1.6 solar masses, and, using the MESA stellar evolutionary code \citep{Paxton2011}, we calculated models of the internal structure, matching as close as possible the stellar parameters determined by grid fitting methods. Specifically, we chose the stellar mass, chemical composition, and the mixing parameter from the grid pipeline YMCM \citep{SilvaAguirre2015} and evolved starting from a pre-main sequence phase to the age estimated from the grid fitted models. Basic parameters of the calculated models are shown in Table~\ref{Table1}.     
\begin{figure}[h]\centering
	\includegraphics[width=\linewidth]{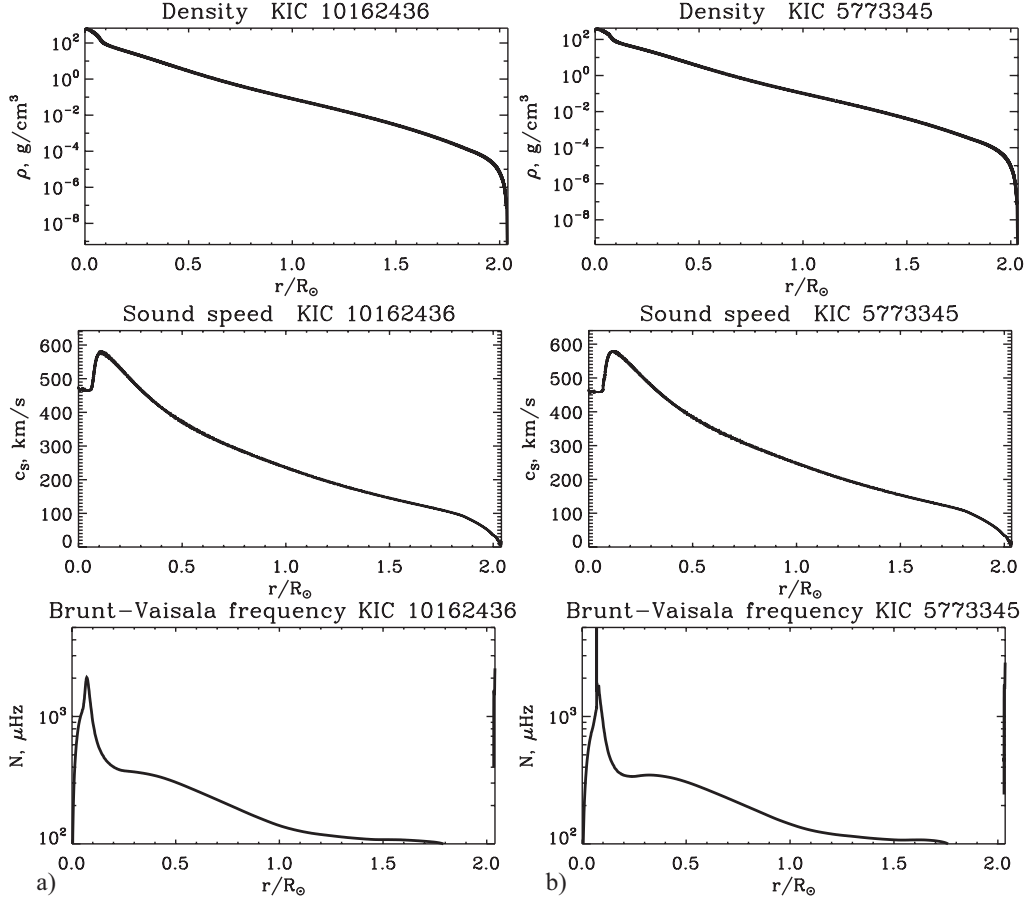}
	\caption{Distributions of the density, the sound speed and the Brunt-V\"ais\"al\"a frequency in the stellar models of: a) KIC 10162436, and b) KIC 5773345.}\label{fig1}
\end{figure}

In this paper, we present results for two models: KIC~10162436 and KIC~5773345, the radial profiles of the density, sound speed, and Brunt-V\"ais\"ail\"a frequency are shown in Figure~\ref{fig1}. The Brunt-V\"ais\"ail\"a frequency displays a sharp peak at the helium core outer boundary, located at about $0.05~R_\odot$ in KIC~10162436, and at $0.07~R_\odot$ in  KIC~577334. 

\begin{figure}\label{fig2}\centering
	\includegraphics[width=\linewidth]{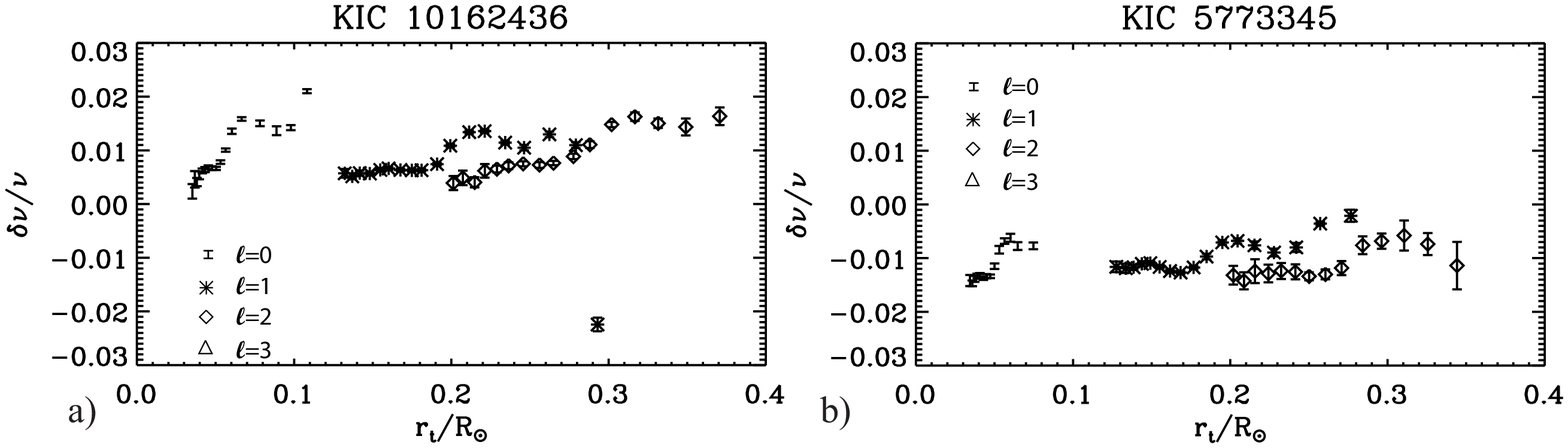}
	\caption{The differences between the observed and modeled frequencies as a function of the acoustic inner turning points for: a) KIC~10162436 and b)KIC~5773345.}
\end{figure}

Figure~\ref{fig2} shows the difference between the observed and modeled frequencies plotted as a function of the radius of the acoustic inner turning points, $r_t$, calculated from the asymptotic relation: $r_t/c_S(r_t)=(l+1/2)/\omega_{nl}$, where $c_S(r)$ is the sound-speed profile, $\ell$ is the angular degree, and $\omega_{nl}$ is the mode frequency.  The acoustic turning points form three branches corresponding to the modes of angular degree $\l=0, 1,$ and 2. Behavior of the frequency deviations for both models is similar, indicating a systematic difference of the evolutionary models from the real stellar structure.

\section{Sensitivity kernels}
The acoustic turning points of some $\ell=0$ modes are located in the stellar core. Nevertheless, the sensitivity of these modes to the core structure is low. Some of the observed low-frequency non-radial modes of $\ell=1$ and 2 represent mixed acoustic-gravity modes which have properties of internal gravity (g) modes in the stellar cores and acoustic (p) modes outside the core. The oscillation frequencies of these modes are very sensitive to the core properties.     
\begin{figure}\centering
	\includegraphics[width=\linewidth]{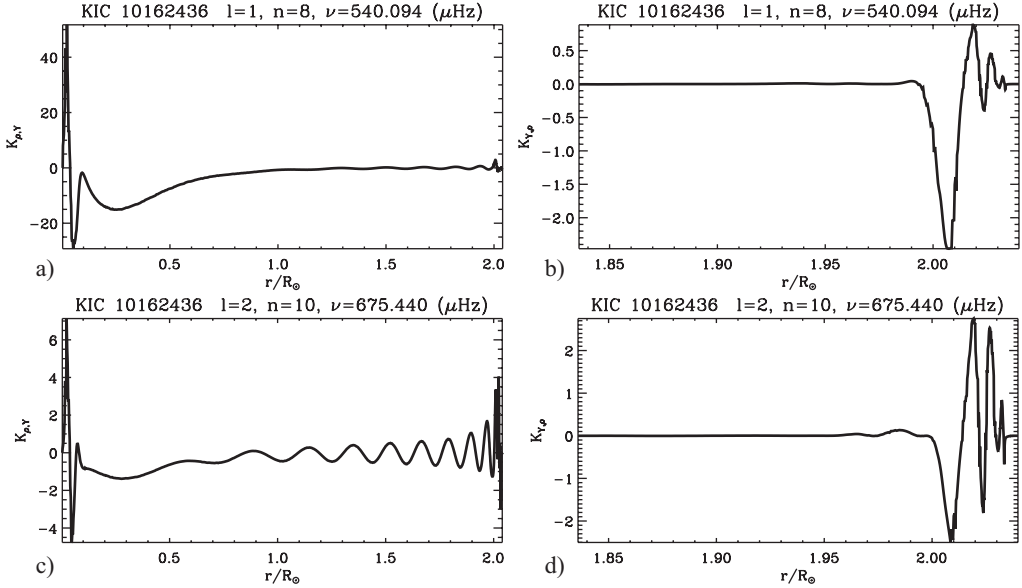}
	\caption{Sensitivity kernels of oscillation frequencies for two mixed modes of $\ell=1$ (panels $a-b$) and $\ell=2$ ($c-d$) to relative variations in stellar density ($a-c$), and variations in the abundance of helium ($b-d$), for  KIC~10162436 ) }\label{fig3}
\end{figure}

 Using explicit formulations for the variational principle, frequency perturbations can be reduced to a system of integral equations for a chosen pair of independent variables; e.g. for $(\rho ,\gamma )$: 
\[\frac{\delta \omega ^{(n,l)} }{\omega ^{(n,l)} } =\int _{0}^{R} \, K_{\rho ,\gamma }^{(n,l)} \frac{\delta \rho }{\rho } dr+\int _{0}^{R} \, K_{\gamma ,\rho }^{(n,l)} \frac{\delta \gamma }{\gamma } dr,\] 
where $K_{\rho ,\gamma }^{(n,l)} (r)$ and $K_{\gamma ,\rho }^{(n,l)} (r)$ are sensitivity (or `seismic') kernels. These are calculated using the initial solar model parameters, $\rho _{0} $, $P_{0} $, $\gamma $, and the oscillation eigenfunctions for these model, $\vec{\xi }$ \citep[for an explicit formulation, see e.g.][]{Kosovichev1999}.
The sensitivity for various pairs of solar parameters, such the sound speed, Brunt-V\"ais\"al\"a frequency, temperature, and chemical abundances, can be obtained by using the relations among these parameters, which follow from the equations of solar structure (`stellar evolution theory'). These `secondary' kernels are then used for direct inversion of the various parameters \citep{Gough1988}. A general procedure for calculating the sensitivity kernels can be illustrated in operator form \citep{Kosovichev2011}. Consider two pairs of solar variables, $\vec{X}$ and $\vec{Y}$, e.g. 
\[\vec{X}=\left(\frac{\delta \rho }{\rho } ,\frac{\delta \gamma }{\gamma } \right);\; \; \vec{X}=\left(\frac{\delta \rho}{\rho} ,\frac{\delta Y}{Y} \right),\] 
where $Y$ is the helium abundance.  The linearized structure equations (the hydrostatic equilibrium equation and the equation of state) that relate these variables can be written symbolically as 
\[A\vec{X}=\vec{Y}.\] 
Let ${\mathop{\vec{K}}\nolimits_{X}} $ and ${\mathop{\vec{K}}\nolimits_{Y}} $ be the sensitivity kernels for $X$ and $Y$; then the frequency perturbation is: 
\[\frac{\delta \omega }{\omega } =\int _{0}^{R} \vec{K}_{X} \cdot \vec{X}dr\equiv \left\langle \vec{K}_{X} \cdot \vec{X}\right\rangle ,\] 
where $<\cdot >$ denotes the inner product. Similarly, 
\[\frac{\delta \omega }{\omega } =\left\langle \vec{K}_{Y} \cdot \vec{Y}\right\rangle .\] 
Then from the stellar structure equation $A\vec{X}=\vec{Y}$: 
\[\left\langle \vec{K}_{Y} \cdot \vec{Y}\right\rangle =\left\langle \vec{K}_{Y} \cdot A\vec{X}\right\rangle =\left\langle A^{*} \vec{K}_{Y} \cdot \vec{X}\right\rangle ,\] 
where $A^{*} $ is an adjoint operator. Thus:                                         $\left\langle A^{*} \vec{K}_{Y} \cdot \vec{X}\right\rangle =\left\langle \vec{K}_{X} \cdot \vec{X}\right\rangle.$
This is valid for any $\vec{X}$ if $A^{*} {\mathop{\vec{K}_{Y} }\nolimits_{}} =\vec{K}_{X}.$ This means that the equation for the sensitivity kernels is adjoint to the stellar structure equations. An explicit formulation in terms of the stellar structure parameters and mode eigenfunctions was given by \citet{Kosovichev1999}.
\begin{figure}\centering
	\includegraphics[width=\linewidth]{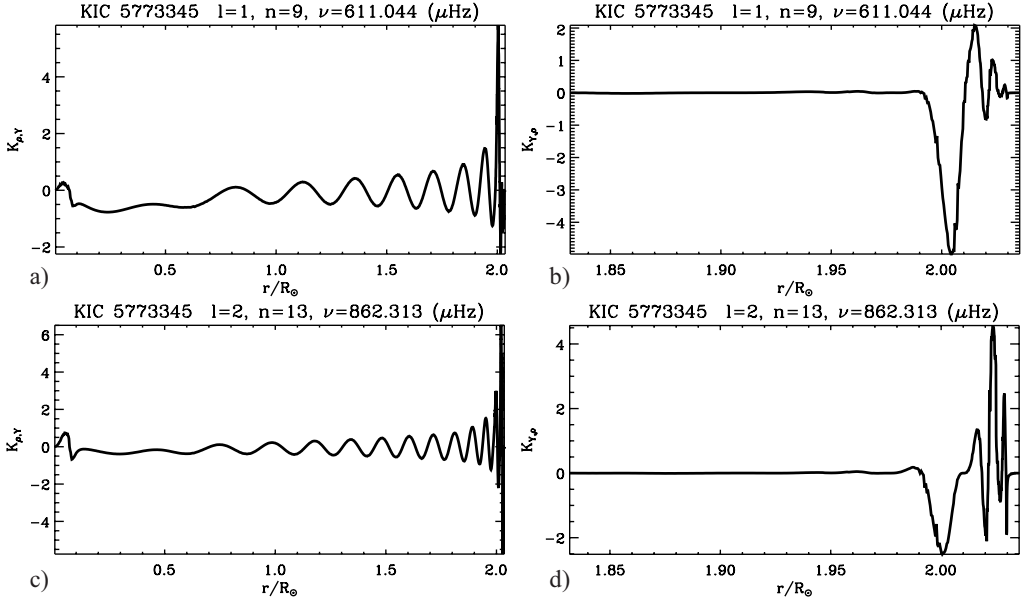}
	\caption{The same as Fig.~\ref{fig3} for  KIC~5773345.}\label{fig4}
\end{figure}

Examples of the sensitivity kernels to density and helium-abundance deviations for the mixed acoustic-gravity modes of $\ell=1$ and $\ell=2$ for KIC 10162436, and KIC 5773345 are shown in Figures~\ref{fig3} and \ref{fig4}, respectively. In the case of KIC 10162436, the sensitivity of the mode frequencies to the density stratification is high in the central core. In the case of KIC 5773345, the sensitivity of the observed mixed modes in the core is not that high but still quite significant.  In such cases the mixed oscillation modes open a unique opportunity for probing the stellar cores by direct structure inversion. The sensitivity kernels for helium abundance are concentrated in the helium and hydrogen ionization zones.

\section{Inversion procedure}

In the inversion procedure it is important to take into account potential systematic uncertainties in the stellar mass and radius. Because the oscillation frequencies are scaled linearly with the factor $q=M/R^3$, then, following \citep{Gough1993}, the mode frequencies $\omega_i$ can be expressed in terms of their
relative small difference $\delta \omega_i^2/\omega_i^2$
from those of a standard reference model of similar mass
and radius according to the linearized expression
\[
\delta \omega_i^2/\omega_i^2 = \int_0^1 \left( K_{f,Y}^i \frac{
	\delta f}{f} + K_{Y,f}^i \delta Y \right) dx - I_q^i \delta q,
\]
where $x = r/R$, $q = M/R^3$, $Y$ is the helium abundance,
and $f$ can be any function of $p$ and $\rho$ \citep{Dappen1991}; 
 $M$ and $R$ are stellar mass and radius,
in solar units, $K_{f,Y}$ and $K_{Y,f}$ are appropriate 
kernels, and $I_q$ is an integral over the reference model.
In this paper we consider two cases: $f=\rho$, and $f=u\equiv P/\rho$ \citep{Dziembowski1991}.
 
These constraints can provide localized averages of $\delta \ln f$
and estimates of $\delta Y$ and $\delta q$ of the kind
\begin{eqnarray*}
	\overline {\delta \ln f} \equiv \int_0^1 \sum_i a_i(x_0)K_{f,Y}^i
	\delta \ln f dx \equiv \int_0^1 A_{f,Y} (x,x_0) \delta \ln f dx
	= 
	\sum_i a_i(x_0)\frac{\delta \omega_i^2}{\omega_i^2}
\end{eqnarray*}
by minimizing over the coefficients $a_i(x_0)$ the functional \citep{Backus1968}:
\begin{eqnarray*},
	\int_0^1 A_{f,Y}^2 (x,x_0) J_f dx + \lambda_1 \int_0^1 \left(
	\sum_i a_i K_{Y,f}^i \right)^2 J_Y dx +
	\lambda_2 \left(\sum_i a_i I_q^i \right)^2 
	+ \alpha \sum_i a_i^2 
	\epsilon_i^2
\end{eqnarray*}
for tradeoff parameters $\lambda_1$, $\lambda_2$ and $\alpha$,
where $J_f=(x-x_0)^2$, and $\epsilon_i$ are standard relative errors in the data.
The tradeoff parameters are chosen empirically, by selecting 
a sufficiently smooth solution and using the L-curve criterion
\citep{Hansen1992}.
\begin{figure}\centering
	\includegraphics[width=\linewidth]{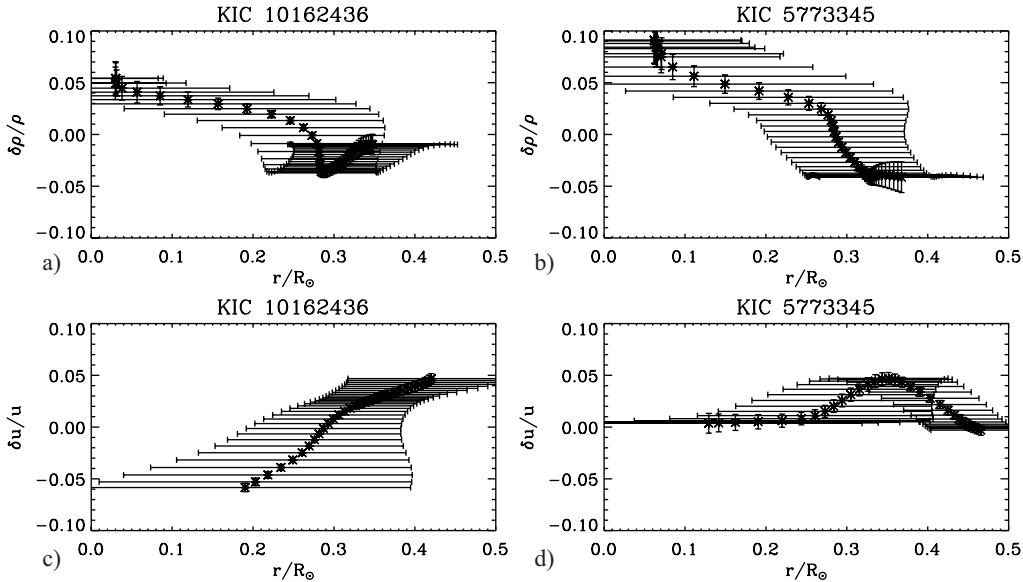}
	\caption{The relative deviations of the stellar structure from the evolutionary models of  density ($a-b$)and  parameter $u=P/\rho$ ($c-d$) as a function of radius for  KIC~10162436 and KIC~5773345, obtained by inversion of the observed mode frequencies. Crosses show the center locations the localized averaging kernels, the horizontal bars show the spread of the averaging kernels, and the vertical bars show the uncertainties calculated using error estimates of the observed frequencies.  }\label{fig5}
\end{figure}

The non-adiabatic frequency shift (aka `surface effect') can be approximated by a smooth function of frequency, $F(\omega)$ scaled with the factor, $Q\equiv I(\omega )/I_{0} (\omega )$, where $I(\omega )$ is the mode inertia, and $I_{0} (\omega )$ is the mode inertia of radial modes ($l=0$), calculated at frequency $\omega $; that is: 
\[\frac{\delta \omega _{{\rm n}onad,i}^{2} }{\omega _{i}^{2} } =F(\omega _{i} )/Q(\omega _{i} ).\] 
Function $F(\omega )$ can be approximated by a polynomial function of degree $K$ \citep{Dziembowski1990}: 
\[F(\omega _{i} )=\sum _{k=0}^{K} c_{k} P_k(\omega _{i}), \]
where $P_k$ are the Legendre polynomials of degree $k$. 
Then the influence of nonadiabatic effects is reduced by applying $K+1$ additional constraints for $a_{i} $: 
\[\sum _{i=1}^{N} a_{i} P_k(\omega _{i}) Q(\omega _{i} )=0,\; \; \; \; k=0,...,K.\] 

\section{Inversion results}
The inversion results obtained by the Optimally Localized Averaging (OLA) procedure described in the previous section are shown in Figure~\ref{fig5}. They show the relative deviations in stellar structure from the evolutionary models of density and parameter $u=P/\rho$ as a function of radius for  KIC~10162436 and KIC~5773345. Crosses show the center locations the localized averaging kernels, the horizontal bars show the spread of the averaging kernels, and the vertical bars show the uncertainties calculated using the observational error estimates \citep[for the definitions of these properties see, ][]{Kosovichev1999}. It is remarkable that the mixed modes observed in the oscillation spectra of these stars allow us to obtain constraints on the density stratification in the central helium core and the hydrogen-burning shell. The averaging kernels for the parameter $u$ (or, equivalently, the sound speed) are localized only outside the core. 
Therefore, it is very important to choose the pairs of inversion variables that provide the best resolution. These variables can be different for different regions of a star. In our cases only the density inversions provide the optimally localized averaging kernels centered in the stellar cores. The sound-speed inversions are incapable of resolving the core. 

For both stars, the inversion results show that the density of the core and the surrounding shell are about 5\% higher than in the stellar models, but lower outside the energy-release shell. The boundary of the helium core is located at $0.05~R_\odot$ in KIC~10162436, and at $0.07~R_\odot$ in  KIC~5773345. Outside the helium cores, the nuclear energy production shells extend $0.3~R_\odot$, with the peak rate at $\sim 0.08~R_\odot$ in both models. Perhaps these stellar regions involve physical processes that are not described by the evolutionary models. For understanding these deviations it will be beneficial to perform more detailed structure inversion studies for a large sample of post-main sequence stars with hydrogen-burning shells.

\section{Conclusion}
High accuracy measurements of oscillation frequencies for a large number stars as well as identification of the observed oscillations in terms of normal modes for stellar models by the grid-fitting techniques open possibilities for performing asteroseismic inversions and for testing evolutionary models. The discovery of the oscillation modes with mixed g- and p-mode characteristics in post-main sequence stars allows us to reconstruct the density stratification in the stellar helium cores and hydrogen burning shell. 

We applied the previously developed asteroseismic inversion method \citep{Gough1993} to two F-type stars from the Kepler Asteroseismic Legacy Project \citep{SilvaAguirre2017a} and performed inversions for the density and squared isothermal sound-speed parameter. The background models were calculated using the MESA code, and the mass, composition, and age were previously determined by the model-grid fitting method. The inversion technique takes potential discrepancies in the estimated mass and radius from the actual properties, as well as the potential frequency shifts due non-adiabatic near-surface effects, and constructs optimally localized averaging kernels following the \citep{Backus1968} method.

The inversion results for both stars showed that the density in the helium core and the inner part of the hydrogen shell may be about 5\% higher than in the evolutionary models, and, in the outer part of the shell, lower. This suggests that the differences may be due to physical processes not described by the evolutionary models. However, more detailed inversion studies for a large sample of stars are needed for quantifying the deviations more precisely.

Acknowledgments. 
The work was partially supported by the NASA Astrophysics Theory Program and grants: NNX14AB7CG and NNX17AE76A.


\end{document}